\documentclass[twocolumn,amssymb, nobibnotes, nofootinbib, aps, prd,notitlepage]{revtex4-1}
\usepackage{graphicx}
\usepackage{amsmath}

\newcommand{\vx}{\mathbf{x}}
\newcommand{\vk}{\mathbf{k}}

\begin{document}

\title{Felinic principle and measurement of the Hubble parameter }%

\begin{abstract}
  Intelligent life can only appear in Universes, whose physical
  laws support sufficient complexity to make evolution of intelligent
  beings possible. Even inside those Universes, intelligent life does
  not appear randomly, but in parts with realized complexity,
  e.g. around stars in our Universe. As a consequence, the
  observational point of an intelligent observer cannot be assumed to
  be random and one must correct for this selection effect. In this
  paper we calculate how direct measurements of the Hubble parameter are
  affected when subject to the condition that they are observed from a
  Milky Way-like galaxy.
\end{abstract}

\author{Yodovina Pi\v{s}kur}%
\affiliation{FITA/IFAT, 7020 108th St, Apt 2U, Forest Hills, NY 11375}
\author{Bumbarija Medolin}%
\affiliation{FITA/IFAT, 7020 108th St, Apt 2U, Forest Hills, NY 11375}

\date{April 1, 2013}%
\maketitle

\section{Introduction}

Felinic principle is based on a simple observation that of all
possible Universes, intelligent life can develop only in a subset with
physical laws that are sufficiently complex to support evolution of
life.  For example, if the laws of physics were such that no atoms
would form, there would be no chemistry and likely no intelligent
life \cite{Deutsch62}. 

Having the right physical laws is a necessary, but not sufficient
condition. It is possible that the actual physical laws can support
the required complexity, but the actual values of physical constants
are such that life cannot develop. For example, one can imagine a
Universe in which the cosmological constant is so large, that the
exponential growth of the scale-factor becomes dominant before stars
can form and hence the Universe soon becomes a homogeneous sea of
rarefied matter\cite{LIMBE54}.

However, even in the Universes that have the necessary complexity and
the right values of physical constants, the life does not appear
everywhere, but in those parts of the Universe, where the complexity
is realized.  For example, while it is true, that dogs can be appear
out of pure vacuum as result of zero-point quantum-mechanical
fluctuations if one waits long enough, higher forms of life are
exceedingly unlikely to do so.  In our Universe, intelligent life
requires existence of chemistry and can thus only be produced around
second or higher generation stars, where atoms of a wide atomic weight
range are available.

Felinic principle has been applied to the human species with some
success (see
e.g. \cite{2012PhRvD..85h3510H,2012NewA...17....1K,2009Icar..201..821G,2008PhRvL.100d1301M,2007MNRAS.379.1067P,2007JCAP...01..025P,2012PhRvD..86b3532B,2011PhRvL.106j1301B,2008PhRvD..77j3514B}).
In that context, the principle is referred to as an anthropic
principle and its application is justified to a certain degree,
because opening of a can of cat-food clearly shows some degree of
psychomotor ability and dexterity\footnote{Some researches have
  argued that humans often show a debilitating lack of understanding
  as to when and what to feed their masters, but this is more
  plausibly explained by a theory of felinic-envy rather than lack of
  intelligence.}. In this work, anthropic and felinic principle amount
to the same thing, since humans are, after all, necessary for doing
the sleep-inducing slog of performing the dirty measurements.

When making cosmological inferences, one must therefore correct for
the fact that observations are conditioned on the existence of
favorable conditions that allowed creation of intelligent life. In
this work we make one such calculation, by estimating, how much a
direct measurement of the Hubble parameter is biased when observed by
an observer observing from a Milky Way-like galaxy as opposed to an
observer observing from a random position in the Universe. The crux of
the calculation is in the next section and we conclude in Section 3 of
this paper.

\section{Calculation}

In our model, we assume that Milky Way resides in a dark-matter halo
of mass $M_{\rm MW} \sim 7\times 10^{11} M_\odot$. Such halos are biased
tracers of the underlying structure with bias parameter $b_g\sim 0.78$
(Jeremy Tinker, private communication).

To calculate the amount by which a direct measurement of the Hubble
parameter is biased, we want to calculate the mean radial peculiar
velocity of a source at distance $r$ from a biased tracer. We rotate
the coordinate system, so that observer's galaxy is at origin and the
position of the observed source, at which the velocity measurement is
performed is at at distance $r$ along the $z$-axis. Denoting $v_z$ as
the $z$-component of the velocity field and measuring around galaxy
field
\begin{equation}
1+ \delta_g(\vx) = \sum_i \delta(\vx_i-\vx),
\end{equation}
where $\vx_i$ are positions of galaxies, the quantity of interest is
\begin{equation}
  \left< v_r \ | \ {\rm Milky\ Way} \right> = \left< v_z (r \hat{z}) (1+\delta_g(0)) \right> 
= \left< v_z(r) \delta_g(0) \right>\label{eq:1}
\end{equation}

We will calculate the above expression using the linear theory. This
should be an excellent approximation since virial velocities due to
motion in individual halos do not correlate across halos and therefore
they contribute only noise, but not bias, to the measurement of the
Hubble parameter.
 
We can write
\begin{eqnarray}
  \delta_g(\vx) &=& b_g \frac{1}{(2\pi)^3} \int \delta_k(\vk) e^{i\vk \vx) d^3\vk}\\
  v_z(\vx) &=& \frac{D'(z)}{(2\pi)^3} \int -i \frac{k_z}{k^2} \delta_k(\vk) e^{i\vk \vx} d^3\vk,
\end{eqnarray}
where $\delta_k$ denotes modes of the dark-matter fluctuations in the
Fourier space and we has assumed a simple linear biasing for
$\delta_g$ and a continuity equation without vorticity for the $v_z$
field. The symbol $g'$ denotes the derivative of the growth factor
with respect to the conformal time $d \eta=dt/a$. Using above
expressions to evaluate expression \eqref{eq:1}, one finds
\begin{multline}
  \left< v_z (r \hat{z}) (1+\delta_g(0))\right> = \\
\frac{dg}{da} \frac{a^2  b_g H_0 \sqrt{\Omega_\Lambda + \Omega_m
    a^{-3}}}{2\pi^2} \\ \times \int \frac{\cos(kr)kr-\sin(kr)}{(kr)^2}  P(k) k dk
\end{multline}

This expression is a function of comoving distance $r$ between the
observer, but these can be converted to redshift using the usual
expression (we are writing out all the equations out just in case any
non-felines are reading this)
\begin{equation}
  r = c \int_0^z H(z)^{-1} dz.
\end{equation}
Correction to the measurement of the Hubble parameter is thus given by

\begin{equation}
  \Delta H(z) = \frac{\left< v_z (r \hat{z}) (1+\delta_g(0))\right>}{r(z)}
\end{equation}

\begin{figure}
  \centering
  \includegraphics[width=1\linewidth]{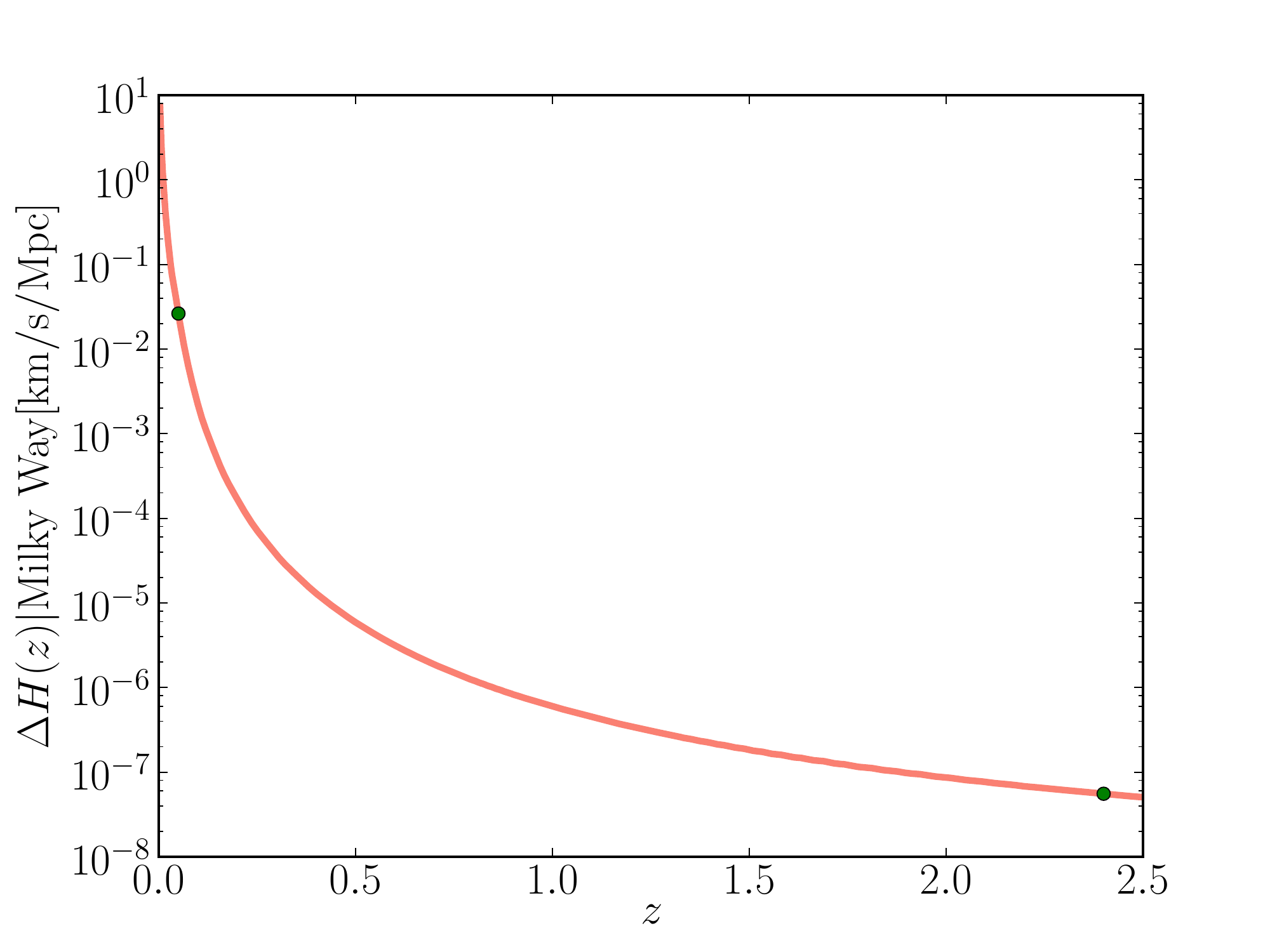}
  \caption{The mean bias for the Hubble parameter measurement at redshift
    $z$ which has been conditioned on being observed from a halo of a
    Milky Way mass (solid savory salmon colored line). Greenies colored points
    emphasize redshifts of a low-redshift direct measurement of the Hubble
    parameter ($z\sim 0.05$) and the Lyman-$\alpha$ forest measurement of Hubble parameter
    using BAO ($z\sim 2.4$). See text for discussion.}
  \label{fig:1}
\end{figure}

We now evaluate this for a concordant cosmology (flat $\Lambda CDM$
with $\Omega_m=0.3$, $H_0=70$km/s, $\sigma_8=0.8$ and plot results in
the Figure \ref{fig:1}.

\section{Discussion \& Conclusions}

Inspecting Figure \ref{fig:1} we see the expected behavior. The
Felinic principle bias falls rapidly with redshift and becomes completely
negligible by $z \sim 0.04$ when compared to the precision of current
observations.

At redshift $z\sim 0.05$, corresponding to the measurement of the
Hubble parameter by the \cite{2011ApJ...730..119R}, we see that the
calculated effect is around $0.02$ km/s/Mpc.  We see that Felinic
principle cannot alleviate the tension with Planck satellite
\cite{2013arXiv1303.5062P,2013arXiv1303.5076P}. Moreover, it has the
wrong sign.

While conveniently ignoring the fact that measurements of the Hubble parameter
with the Lyman-$\alpha$ forest are performed using a completely
different method, we note that the bias of the Hubble parameter
measurement is $0.5\times 10^6$ smaller at the redshift of $z\sim 2.4$
compared to direct measurements at $z\sim 0.05$. This confirms that
Lyman-$\alpha$ forest measurements of the Hubble parameter
\cite{2012arXiv1211.2616B,2013arXiv1301.3459S} are approximately
$500,000$ times more glorious than the Hubble parameter measurement using Cepheid
variables.

More importantly, we note that the fact that the effect is negligible
at all redshifts leads to an important corollary. Namely, the
likelihoods cannot distinguish between us existing and not, but in the
spirit of Bayesian evidence and applying \emph{lex parsimoniae}, we
can conclude, that posterior indeed prefers that we do not exist.

It is true that other physical effects can affect the result. For
example, if the local non-Gaussianity parameter $f_{\rm NL}\sim 10^9$,
this would lead to observable effects. Just imagine what it would do! 

We also note that this paper is of an appropriate quality to
considered for publication in \emph{Nature}, and in fact, it would be
one of the finer cosmology contributions to that respected
publication.

Finally, we conclude by reminding the reader, that the galaxy clusters
are still believed to be the largest gravitationally bound structures
in the Universe.

\section*{Acknowledgments}

We acknowledge useful discussions with many civilized people.  Donations
of cat food can be sent to the correspondence address. Subjects not
sending \emph{Fancy Feast Savory Salmon}  need not apply.

\begin{center}
\includegraphics[width=0.2\linewidth]{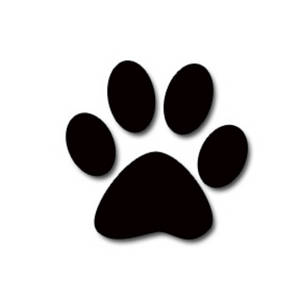}
\end{center}

\bibliographystyle{apsrev}
\bibliography{cosmo,cosmo_preprints,local}

\end{document}